\documentclass[aps,showpacs,11pt]{revtex4}
\usepackage{dcolumn}
\usepackage{graphicx}
\usepackage{amsmath}
\usepackage{amsfonts}
\usepackage{amssymb}
\usepackage{psfrag}
\usepackage{wrapfig}
\usepackage{subfigure}
\usepackage{makeidx}
\usepackage{bm}
\usepackage{epsf}
\usepackage{hyperref}
\makeatletter

\newcommand{\Rmnum}[1]{\expandafter\@slowromancap\romannumeral #1@}
\makeatother

\begin{document}

\title{Charged black holes in the Einstein-Maxwell-Weyl gravity}

\author{Chao Wu$^{1}$\footnote{wc130330@163.com},\ De-Cheng Zou$^{1}$\footnote{Corresponding author: dczou@yzu.edu.cn},\ Ming Zhang$^{2}$\footnote{zhangming@xaau.edu.cn}}
\affiliation{
$^{1}$Center for Gravitation and Cosmology and College of Physical Science and Technology,\\ Yangzhou University, Yangzhou 225009, China\\
$^{2}$Faculty of Science, Xi'an Aeronautical University, Xi'an 710077 China
}

\date{\today}

\begin{abstract}
\indent

We construct charged asymptotically flat black hole solutions in
Einstein-Maxwell-Weyl(EMW) gravity. These solutions can be interpreted as
generalizations of two different groups: Schwarzschild black hole (SBH)
and non-Schwarzschild black hole (NSBH) solutions. In addition,
we discuss the thermodynamic properties of two groups
of numerical solutions in detail, and show that they
obey the first law of thermodynamics.

\end{abstract}



\maketitle


\section{Introduction}
\label{1s}

As well known, the general relativity can be viewed as an effective
low-energy field theory  in string theory. It's also the first term
in an infinite series of gravitational corrections
built from powers of the curvature tensor and its derivatives~\cite{tHooft:1974toh}.
In addition, the general relativity leaves some open fundamental questions,
including the problems of singularity, non-renormalizability, the dark matter
and dark energy. In order to answer these questions, Stelle et.al asserted
one can add higher derivatives terms to the Einstein-Hilbert action~\cite{Stelle:1976gc}.
In four-dimensional spacetime, the most general theory up to the second
order in the curvature is given by~\cite{Lu:2015cqa,Lu:2015psa,Holdom:2016nek}
\begin{eqnarray}
{\cal I}=\int d^{4}x \sqrt{-g}\left[\gamma R-\alpha C_{\mu\nu\rho\sigma}C^{\mu\nu\rho\sigma}
+\beta R^2\right]\label{action0}
\end{eqnarray}
where $\alpha$, $\beta$ and $\gamma$ are constants, and $C_{\mu\nu\rho\sigma}$ is the Weyl tensor.
Notice that for any static, spherically symmetric black-hole solution,
the term quadratic in Ricci scalar $R$ makes no contribution to the
corresponding field equations \cite{Nelson:2010ig}.
As a result, this setting of $\beta=0$ and $\gamma=1$ has been often applied
for simplicity in the Einstein-Weyl (EW) gravity, and this theory
reduces to the pure Einstein-Weyl gravity.
By fixed $\alpha=1/2$, L\"{u} et.al~\cite{Lu:2015cqa,Lu:2015psa} derived a new numerical
non-Schwarzschild black hole (NSBH) in the region of $0.876<r_0<1.143$
for horizon radius $r_0$ in the EW gravity,
besides the Schwarzschild black hole (SBH).
Different from SBH, this NSBH has some remarkable properties: i) it admits
positive and negative values of black hole mass; ii)
when the horizon radius $r_0$ approaches some extremal value,
this new black hole approaches the massless state~\cite{Lu:2017kzi}.
In terms of the continued fractions, Kokkotas et.al \cite{Kokkotas:2017zwt}
further constructed this NSBH solution in the analytical form.
Very recently, the new black hole solutions have been also derived by adopting
a new form of the metric in the pure Einstein-Weyl gravity~\cite{Podolsky:2018pfe}.
Under test scalar field perturbation, the stabilities of
NSBH and SBH have been also separately investigated ~\cite{Cai:2015fia,Zinhailo:2018ska},
which recovered that the quasinormal modes for NSBH share larger real
oscillation frequency and larger damping rate than the SBH branch.
In particular, undamped oscillations (called quasi-resonances) also emerged for
the NSBH, if perturbed scalar field possesses sufficiently large
masses.

Inspired by above issues, Lin et.al further explored the Einstein-Maxwell-Weyl(EMW)
gravity consisting of pure EW term and electromagnetic field~\cite{Lin:2016jjl},
where it presented two groups (Groups \textbf{I} and \textbf{II})
of new charged black holes with fixed $\beta=0$, $\gamma=1$ and $\alpha=1/2$.
Actually, the two groups solutions could be separately viewed as a charged
generalization of the higher derivative curvature for SBH
and a charged generalization of NSBH.
However, Reissner-Nordstr$\ddot{o}$m (RN) black hole
solution is not a solution in the EMW gravity.

Comparing with the NSBH in the bound $0.876<r_0<1.143$~\cite{Lu:2015cqa,Lu:2015psa},
new NSBH solutions have been obtained
within an extended bound $0.363<r_0<1.143$ in pure EW gravity\cite{Lu:2017kzi}.
With regard to the EMW gravity, we can also
construct new charged black hole solutions according to the new non-charged
`seed' solutions. Then, we will further discuss the thermodynamic properties
of these new charged black holes in the EMW gravity.

This paper is organized as follows. In Sec.~\ref{2s}, we numerically
derive new charged asymptotically flat black hole
solutions. Then, some related thermodynamic properties
of these new charged black holes are explored in Sec.~\ref{3s}.
Finally, we end the paper with conclusions and discussions in Sec.~\ref{4s}.

\section{Einstein-Maxwell-Weyl gravity and numerical solutions}
\label{2s}

The Einstein-Maxwell-Weyl action including pure Einstein-Weyl term
and electromagnetic field is given by~\cite{Lin:2016jjl}
\begin{eqnarray}
{\cal I}=\int d^{4}x \sqrt{-g}\left[R-\alpha C_{\mu\nu\rho\sigma}C^{\mu\nu\rho\sigma}
-\kappa F_{\mu\nu}F^{\mu\nu}\right]\label{action},
\end{eqnarray}
where $F_{\mu\nu}=\nabla_{\mu}A_{\nu}-\nabla_{\nu}A_{\mu}$ is the electromagnetic tensor.
Then, the equations of motion are obtained as~\cite{Lin:2016jjl}
\begin{eqnarray}
R_{\mu\nu}-\frac{1}{2}g_{\mu\nu}R-4\alpha B_{\mu\nu}-2\kappa T_{\mu\nu}&=&0,\nonumber\\
\nabla_{\mu}F^{\mu\nu}&=&0,\label{eom}
\end{eqnarray}
where the trace-free Bach tensor $B_{\mu\nu}$ and energy-momentum tensor of
electromagnetic field $T_{\mu\nu}$ are defined as
\begin{eqnarray}
B_{\mu\nu}=\left(\nabla^\rho\nabla^\sigma+\frac{1}{2}R^{\rho\sigma}\right)C_{\mu\nu\rho\sigma},\quad
T_{\mu\nu}=F_{\alpha\mu}F^{\alpha}_{~\nu}-\frac{1}{4}g_{\mu\nu}F_{\alpha\beta}F^{\alpha\beta}.
\end{eqnarray}

In order to construct new charged black hole solution, we assume a new metric ansatz
\begin{eqnarray}
ds^2=-f(r)e^{-2\delta(r)}dt^2+\frac{1}{f(r)}dr^2+r^2\left(d\theta^2+\sin^2\theta d\varphi^2\right),\label{nansatz}
\end{eqnarray}
with $h(r)=f(r)e^{-2\delta(r)}$ and $f(r)=1-\frac{2m(r)}{r}$.
Substituting the ansatz [eq.(\ref{nansatz})] into eq.(\ref{eom}), one get
\begin{eqnarray}\label{neom}
&&2r^3m'-Q^2r+2r^3(r-2m)\delta'+\alpha\Big[4r(2m-r)\left(r-3m+r m'+r(r-2m)\delta'\right)\delta''\nonumber\\
&&+16(2m-r)m'+4\left(4r^2m'^2-2r^2+2m^2+5m r+(r^2-10mr)m'\right)\delta'\nonumber\\
&&+\left(2r\left(28m^2-16rm+r^2\right)+24r^2(r-2m)\right)\delta'^2+8r^2(r-2m)^2\delta'^3\Big]=0,\label{neom1}\\
&&2r^3m'-Q^2r+2r^3(r-2m)\delta'+\alpha\Big[4r\left(r-3m+r m'+r(r-2m)\delta'\right)m''\nonumber\\
&&+8(m-r)m'+8m'^2r+4\left(r^2m'^2-2m r+5m^2+2(r^2-3m r)m'\right)\delta'\nonumber\\
&&+\left(2r\left(20m^2-16rm+3r^2\right)+8r^2(r-2m)\right)\delta'^2+4r^2(r-2m)^2\delta'^3\Big]=0,\label{neom2}\\
&&A_t'-\frac{Q e^{-\delta}}{r^2}=0, \label{neom4}
\end{eqnarray}
where the prime ($'$) denotes differentiation with respect to $r$, and parameter $Q$
denotes the electric charge.

Influenced by the functional form of the electric charge in the metric,
the charged black holes have more than one horizon in general. For example,
RN black hole has one event horizon and one Cauchy horizon.
However, we suppose that the spacetime has only one horizon to make it
easier for the expansion of $m(r)$ and $\delta(r)$ around the event horizon $r_0$.
\begin{eqnarray}\label{nexpr}
&&m(r)=\frac{r_0}{2}+m_1(r-r_0)+m_2(r-r_0)^2+\ldots,\label{aps-1}\\
&&\delta(r)=\delta_0+\delta_1(r-r_0)+\delta_2(r-r_0)+\ldots,\label{aps-2}\\
&&A_t(r)=A_{t1}(r-r_0)+A_{t2}(r-r_0)^2+\ldots\label{aps-4} .
\end{eqnarray}
Substituting these expansions into eq.(\ref{eom}), the coefficients $\delta_i$,
$A_{ti}$ (for $i=1$) and $m_i$ (for $i=2$) can be solved in terms of the
three non-trivial free parameters $r_0$, $m_1$ and $\delta_0$.
For example, $m_2$, $A_{t1}$ and $\delta_1$ can be obtained as
\begin{eqnarray}\label{ncoef}
 m_2=-\frac{m_1}{r_0}-\frac{6m_1r_0^2-3Q^2}{16\alpha(2m_1-1)r_0},\quad
\delta_1=\frac{2m_1r_0^2-Q^2}{4\alpha(2m_1-1)^2r_0},\quad A_{t1}=\frac{e^{-\delta_0}Q}{r_0^2}.
\end{eqnarray}
Here the coefficients of expansion can be also presented by $m_1=-\frac{\delta}{2}$, and $\delta_0=\frac{1}{2}\ln(\frac{1+\delta}{c r_0})$ in refs.~\cite{Lu:2015cqa,Lu:2015psa}.

At the other asymptotic regime, that of radial infinity $(r\rightarrow1)$, the metric functions and
the Maxwell field may be expanded in power series, this time in terms of $1/r$.
The metric components reduce to
\begin{eqnarray}\label{ncoef}
m(r)=M-\frac{Q^2}{2r}+\ldots,\quad
\delta(r)=\frac{2\alpha Q^2}{r^4}+\ldots,~A_t(r)=\Phi+\frac{Q}{r}+\ldots, \label{insol}
\end{eqnarray}
where the parameter $M$ and $Q$ are associated with the mass and charge of black hole,
and $\Phi$ is the electric potential.

\begin{figure}[htb]
\centering
\subfigure[$r_0=0.6, m_1=0.313, \delta_0=0.553$]{
\includegraphics{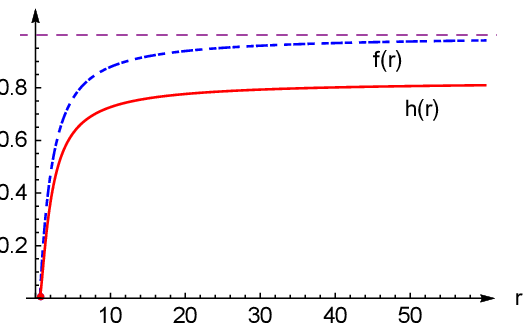}}
\hfill%
\subfigure[$r_0=1, m_1=-0.182, \delta_0=0.011$]{
\includegraphics{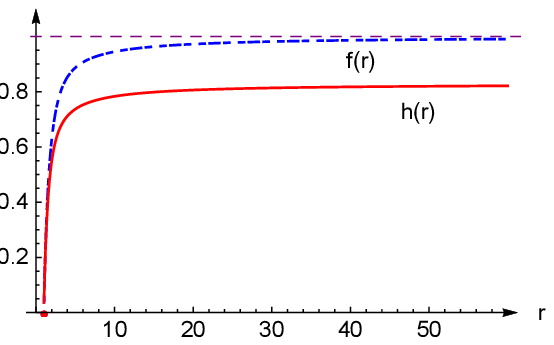}}
\caption{The NSBH solutions for metric functions $f(r)$ and $h(r)$.
For clarity we have chosen a rescaling of $h(r)$ with 4/5, rather than 1,
to avoid an asymptotic overlap of the curves.}\label{fig1}
\end{figure}

Firstly, we reconsider non-Schwarzschild black hole solution $(Q=0)$ in the EW gravity.
Throughout this paper, we also take $\alpha=\frac{1}{2}$ and $\kappa=1$
for simplicity \cite{Lu:2015cqa,Lu:2015psa}.
The signal for a good black hole solution is that the functions $f(r)$
and $h(r)$ should approach very close to 1 as $r$ increases.
Comparing with the previous bound $(0.876<r_0<1.143)$ for the
horizon radius $r_0$ \cite{Lu:2015cqa,Lu:2015psa},
we numerically derive an extended bound $0.363<r_0<1.143$.
The values $0.363$ and $1.143$ of horizon radius $r_0$ denote the
disappearance of temperature $T=0$ and massless state $M=0$
for the non-Schwarzschild black hole, respectively.
If $r_0$ is smaller than 0.363, the temperature of non-Schwarzschild black
hole is negative. When $r_0$ is larger than 1.143, the black hole mass would
become negative \cite{Lu:2015cqa}.
The corresponding bound for the parameter $m_1$ is $0.46>m_1>-0.422$.
At $r_0\approx0.876$, the Schwarzschild and the non-Schwarzschild black
holes 'coalesce' with $m_1=0$.
In other words, for any selected value of $r_0$ in the above extended
bounded interval, there exists only one value of $m_1$ that
allows for a healthy non-Schwarzschild black hole.
In Fig.{\ref{fig1}}, we plot the metric functions $f$ and $h$
for the NSBH with horizon radius $r_0=0.6$ and 1.

\begin{figure}[htb]
\centering
\subfigure[$r_0=0.5,m_1=-0.0194$]{
\includegraphics{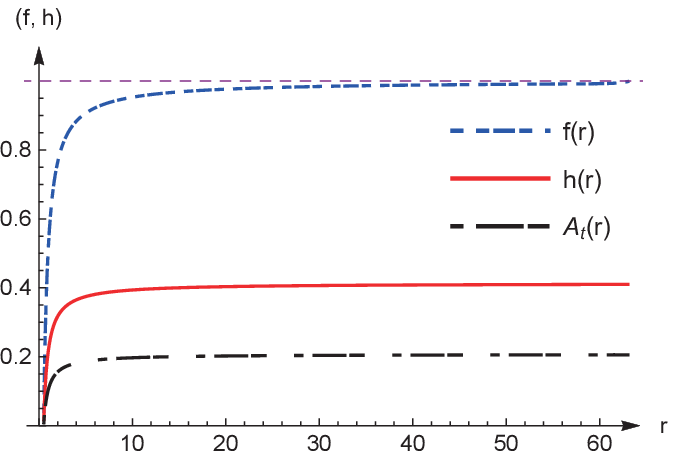}}%
\hfill%
\subfigure[$r_0=2,m_1=0.0037$]{
\includegraphics{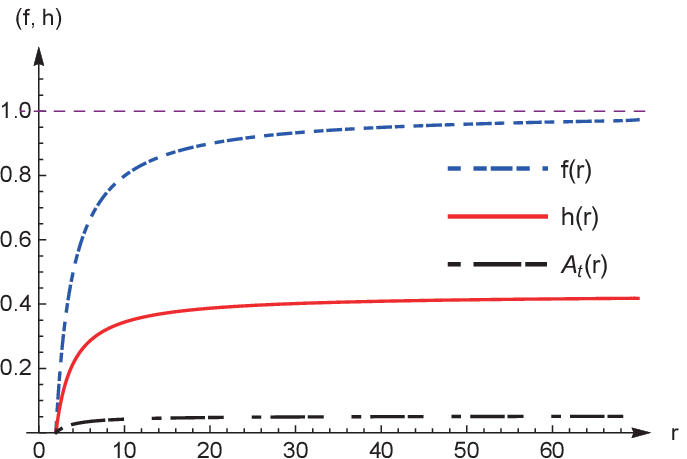}}
\caption{Numerical charged solutions for $f(r)$, $h(r)$ and $A_t(r)$ and
some values of $Q=0.16$ in Group \textbf{I}. In each plot the
function $h(r)$ with $\delta_0=0.423$ is chosen to approach to 4/5
instead of 1 for clarity. The purple dashed
line represents the unity.}\label{fig2}
\end{figure}

\begin{figure}[htb]
\centering
\subfigure[$r_0=0.7,m_1=0.248,\delta_0=0.710$]{
\includegraphics{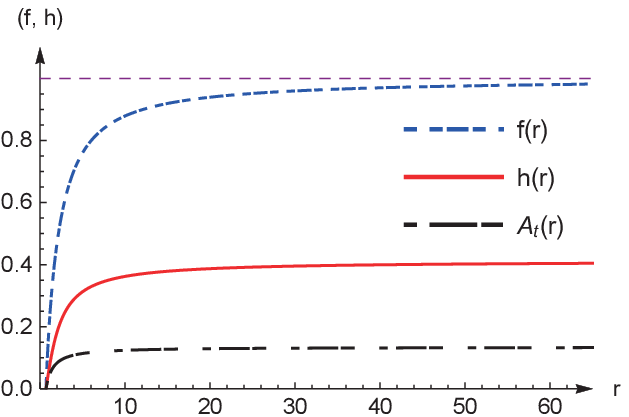}}%
\hfill%
\subfigure[$r_0=1.1,m_1=-0.370,\delta_0=0.328$]{
\includegraphics{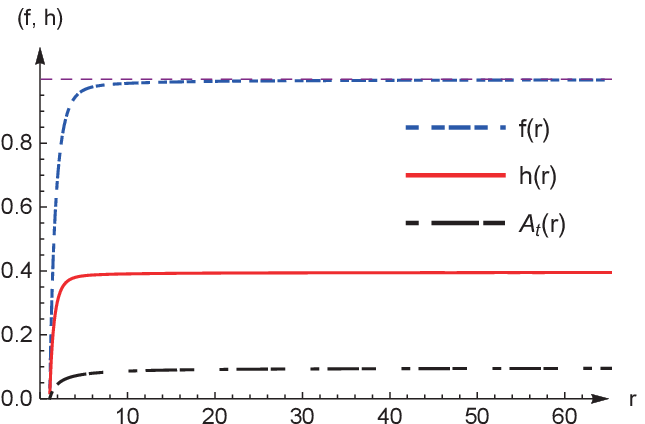}}
\caption{Numerical charged solutions for $f(r)$, $h(r)$ and $A_t(r)$
and $Q=0.16$ in Group \textbf{II}. In each plot the function $h(r)$
is chosen to approach 2/5 for clarity. The purple dashed line
represents the unity.}\label{fig3}
\end{figure}

Now, we turn to discuss the charged black hole in the Einstein-Maxwell-Weyl gravity.
It's worth noticing that the ref.\cite{Lin:2016jjl} asserted  the metric
functions $f(r)$ and $h(r)$ for the charged black hole solutions in
the Group \textbf{II} presented a peak outside the event horizon.
It implied the presence of a unphysical negative effective mass [Fig.3 in ref.~\cite{Lin:2016jjl}].
Here we reconstruct four different charged black hole
solutions of groups \textbf{I} and \textbf{II} on both sides
of the coalesce point $r_0\approx0.876$ of Schwarzschild
and non-Schwarzschild black holes, see Figs.~{\ref{fig2}} and {\ref{fig3}}.
Obviously, the peak of the functions $f(r)$ and $h(r)$ vanishes,
which means that new charged black holes of the Group \textbf{II}
could possess positive masses. We will verify it in the next section.

\section{Thermodynamic properties of charged black holes}
\label{3s}

With the numerical charged black holes of Groups \textbf{I} and \textbf{II},
it's interesting to study the thermodynamic properties of these solutions.
In order to do this, we need to collect the numerical results for a sequence of
charged black-hole solutions with different values of $Q$.

\begin{figure}[htb]
\centering
\subfigure[Group \textbf{I}]{
\includegraphics{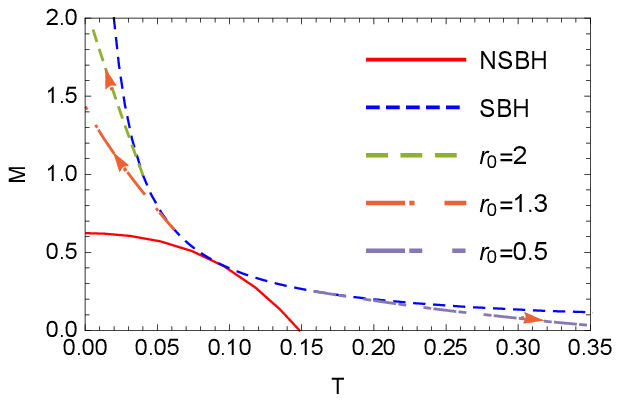}}
\hfill%
\subfigure[Group \textbf{II}]{
\includegraphics[width=0.36\textwidth]{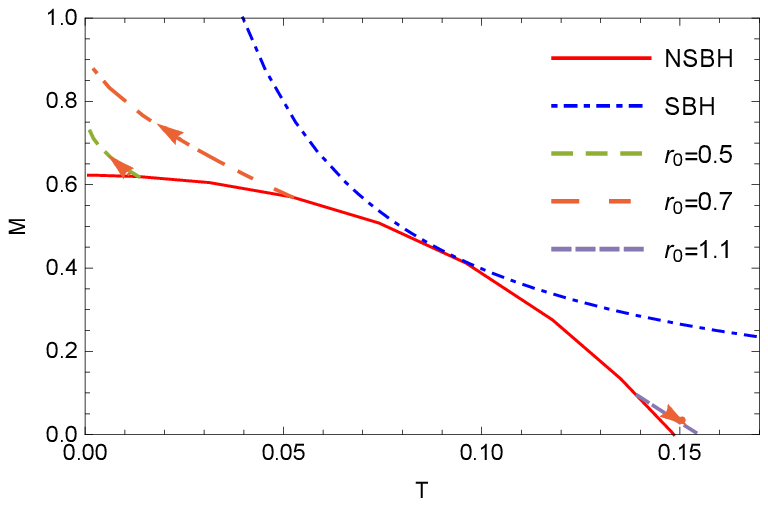}}
\caption{Mass $M$ versus temperature $T$ relations for charged, SBH and NSBH.
The arrow indicates the increase of charge $Q$.}\label{fig4}
\end{figure}

\begin{figure}[htb]
\centering
\subfigure[Group \textbf{I}]{
\includegraphics{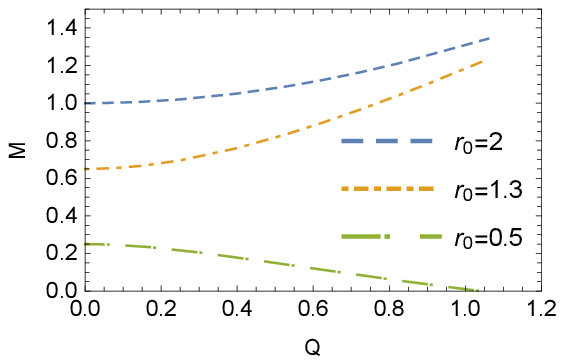}}
\hfill%
\subfigure[Group \textbf{II}]{
\includegraphics{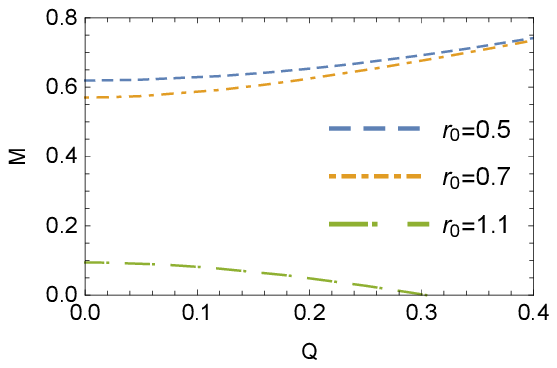}}
\caption{Mass $M$ versus charge $Q$ relations for charged black holes
in the Groups \textbf{I} and \textbf{II}. }\label{fig5}
\end{figure}

In Fig.~{\ref{fig4}}, we show the mass $M$ as a function of Hawking
temperature $T$ for the SBH and NSBH in the pure Einstein-Weyl gravity.
At high temperature region, SBH and NSBH take small masses $M$,
coalesce at $T\approx0.091(r_0\approx0.876)$,
and then take reverse behaviors at lower temperatures.
Take the charged black holes with $r_0=1.3$ of the Group \textbf{I} as an example:
starting from SBH and holding the horizon radius $r_0=2$, the mass
of charged black hole becomes larger, while the temperature decreases
as the increase of charge $Q$, and then vanishes at $Q\approx1.86$,
see Fig.~{\ref{fig4}}(a). Here the arrow indicates the
increase of charge $Q$.  On the other hand, the temperature
increases and mass decreases when the charge $Q$ becomes larger
for the new charged black holes bifurcation from the
right-hand part of coalesce point.
There also exist a bound $0<Q\leq1.035$ for the
charged black hole with $r_0=0.5$.
Interestingly, if $Q$ is increased beyond $Q_c=1.035$,
one can still obtain a charged black hole solution
for an appropriate choice of $m_1$, but now the
mass is actually negative, see Fig.~{\ref{fig5}}(a).

The similar phenomenon occurs for charged black holes in the Group \textbf{II}.
We construct charged black holes starting
from a sequence of non-Schwarzschild black holes with
horizon radius in the region of $0.363<r_0<1.143$.
see Fig.~{\ref{fig4}}(b). Considering $M\geq0$ and $T\geq0$,
the charge $Q$ of charged black holes need to satisfy $0<Q<0.305$ for $r_0=1.1$,
$0<Q<0.4$ for $r_0=0.5$ and $0<Q<0.65$ for $r_0=0.7$, see Fig.~{\ref{fig5}}(b).

\begin{figure}[htb]
\centering
\subfigure[Group \textbf{I}]{
\includegraphics{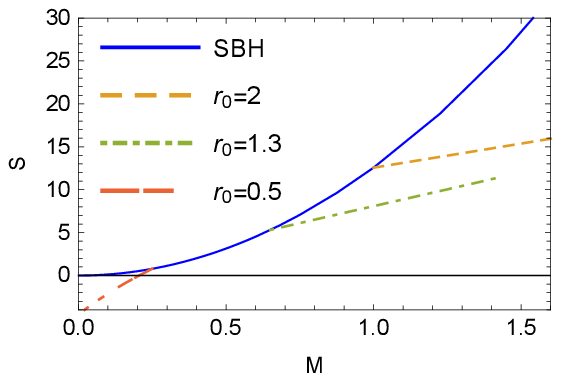}}
\hfill%
\subfigure[Group \textbf{II}]{
\includegraphics{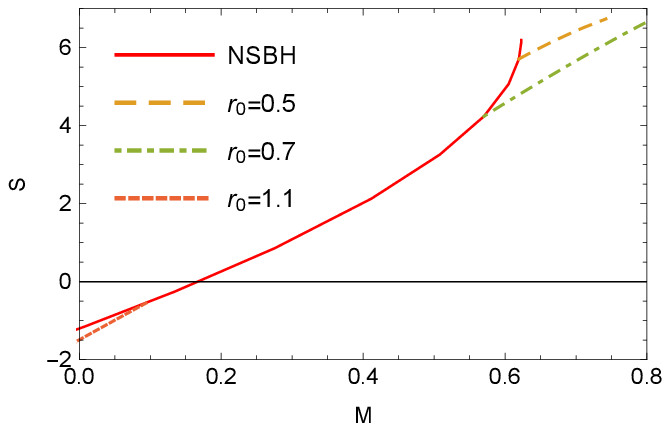}}
\caption{Entropy $S$ versus mass $M$ for the charged black holes. }\label{fig6}
\end{figure}

Due to the higher-derivative theory, the entropy cannot be simply given by
one quarter of the area of the event horizon, which equals to $S=\pi r_0^2+8\pi\alpha m_1$.
The entropy $S$ as a function of mass $M$ of these charged black holes in the
Groups \textbf{I} and \textbf{II} are shown in Figs.~{\ref{fig6}}.

Now we check the first law of thermodynamics of charged black holes by utilizing
the numerical values for thermodynamic quantities $M$, $S$, $Q$, $T$ and $\Phi$.
We only consider the charged black hole with $r_0=0.5$ in the
Group \textbf{I} and \textbf{II}. These discrete values of these
thermodynamical quantities are shown in the TABLE I.
In the table on the left (charged black hole in Group \textbf{I}),
forward differences of mass $M$ entropy $S$ and charge $Q$ can be written as
\begin{eqnarray}
\Delta M&\equiv&\frac{M[i+2]-M[i]}{2},\nonumber\\
&=&\{0.0002635, 0.0013918, 0.0026031, 0.003350, 0.0042297, 0.0060, 0.0068047\},\nonumber\\
\Delta S&\equiv&\frac{S[i+2]-S[i]}{2}, \nonumber\\
&=&\{0.0044782, 0.023697, 0.044434, 0.0574451, 0.07291, 0.104281, 0.119241\}, \nonumber\\
\Delta Q&\equiv&\frac{Q[i+2]-Q[i]}{2}
=\{-0.015, -0.03, -0.035, -0.03, -0.03, -0.035, -0.035\},\quad i=1..7. \label{mvalu}
\end{eqnarray}
Then the expression $dM-\left(TdS+\Phi dQ\right)$ in the form of discrete
points is given by
\begin{eqnarray}\label{firstlaw1}
&&\Delta M[i]-\left(T[i+1]\cdot\Delta S[i]+\Phi[i+1]\cdot\Delta Q[i]\right) \nonumber\\
&&= \{-1.5*10^{-4}, -5.8*10^{-4}, 3.6*10^{-4},
1.6*10^{-5}, 2.0*10^{-5}, -2.6*10^{-4}, 3.2*10^{-4}\},\quad i=1..7.\nonumber\\
\end{eqnarray}
From the Table on the right (charged black hole in Group \textbf{II}), 
we can also calculate $dM-\left(TdS+\Phi dQ\right)$ by using the finite
difference method
\begin{eqnarray}\label{firstlaw2}
&&\Delta M[i]-\left(T[i+1]\cdot\Delta S[i]+\Phi[i+1]\cdot\Delta Q[i]\right) \nonumber\\
&&= \{6.8*10^{-5}, 2.2*10^{-5}, 6.6*10^{-5}, 2.1*10^{-6}, \
6.6*10^{-5}, 1.2*10^{-6}, 6.0*10^{-5}\},\quad i=1..7.\nonumber\\
\end{eqnarray}
Thus the charged black holes obey the first
law $dM=TdS+\Phi dQ$ to quite a high precision.

\begin{table}[h]\label{table1}
{\begin{tabular}{|c|c|c|c|c|c|}
\hline
No. & Q & M & S & T &$\Phi$   \\ \hline
1&0&0.25& 0.78539&0.159155& 0\\ \hline
2&1/100 & 0.24994& 0.78440& 0.15919& 0.0200 \\ \hline
3&3/100& 0.24947& 0.77644& 0.15949& 0.06001  \\ \hline
4&7/100& 0.24715& 0.73700& 0.16097& 0.14023  \\ \hline
5&10/100& 0.24426&0.68757& 0.16283& 0.20066 \\ \hline
6&13/100& 0.24045& 0.62211& 0.16530& 0.26143  \\ \hline
7&16/100& 0.23580& 0.54175& 0.16835& 0.32261 \\ \hline
8&20/100& 0.22845& 0.41355&0.17325& 0.40497\\ \hline
9&23/100& 0.22220&0.30327& 0.17749& 0.46737\\ \hline

\end{tabular}}
\hfill%
{\begin{tabular}{|c|c|c|c|c|c|}
\hline
No. &Q & M & S & T &$\Phi$   \\ \hline
1&0& 0.61944&5.71036&0.01348& 0 \\ \hline
2&2/100& 0.619849&5.71391& 0.013413& 0.031039 \\ \hline
3&35/1000 &0.62054& 5.72123& 0.01328& 0.054268 \\ \hline
4&5/100& 0.62168 & 5.73247& 0.013079& 0.07744 \\ \hline
5&7/100& 0.62380& 5.75344& 0.012705& 0.10821  \\ \hline
6&9/100& 0.62662& 5.78105& 0.012218& 0.13877  \\ \hline
7&105/1000& 0.62917& 5.80592& 0.01178& 0.16152 \\ \hline
8&12/100& 0.63208& 5.8342&0.01129& 0.18409 \\ \hline
9&14/100& 0.63653& 5.87694& 0.010576& 0.21388 \\ \hline
\end{tabular}}
\caption{ The discrete values of thermodynamical quantities $M$, $T$, $S$ and $\Phi$
for charged black holes with $r_+=0.5$ in the Group \textbf{I} (Left) and Group \textbf{II} (Right).}
\end{table}

\begin{figure}[htb]
\centering
\subfigure[Group \textbf{I}]{
\includegraphics{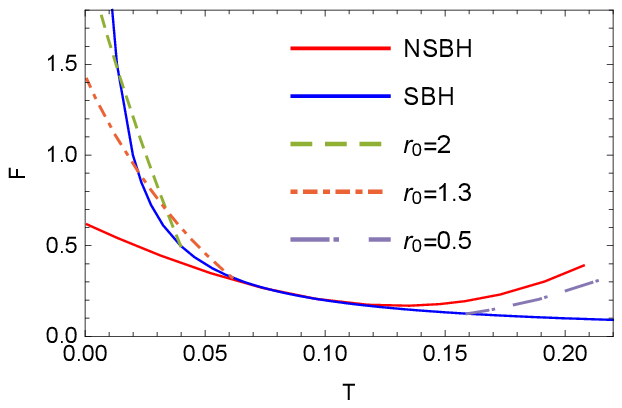}}
\hfill%
\subfigure[Group \textbf{II}]{
\includegraphics{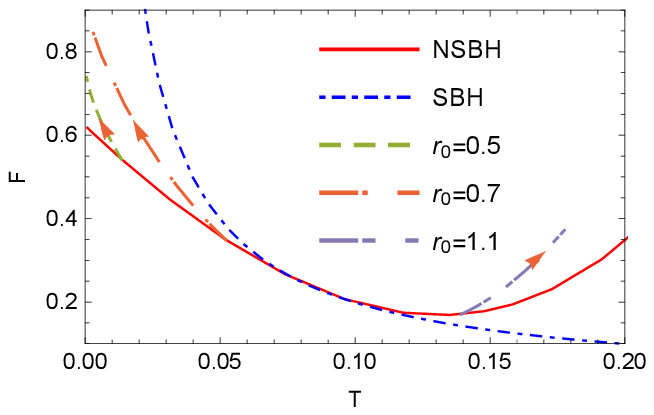}}
\caption{Free energy $F$ versus temperature $T$ relations for the charged,
Schwarzschild and non-Schwarzschild black holes families.}\label{fig7}
\end{figure}

It is important to explore the free energy $F=M-TS$ as a function of temperature.
According to the left-hand side of joint point ($T\approx0.091$) in Fig.{\ref{fig7}}(b),
it can be seen that free energies of the charged black holes
of Group \textbf{II} are always larger than that of non-Schwarzschild black hole, but smaller
than that of Schwarzschild black hole at a given temperature $T$. Nevertheless,
the charged black holes located on the right-hand side of joint point ($T\approx0.091$)
always possess larger values for fixed $T$. The charged black holes in Group \textbf{I} display more complicated properties in Fig.{\ref{fig7}}(a). The free energy of charged black hole for left-hand side of joint point ($T\approx0.091$) is larger than that of Schwarzschild black hole within
some region of $T(>T_c)$, while becomes smaller when the temperature satisfies
$T>T_c$ [$T_c\approx0.023$ for $r_0=1.3$ and $T_c\approx0.017$ for $r_0=2$].

\section{Conclusions and discussions}
\label{4s}

In this paper, we obtained numerically charged asymptotically flat black
hole solutions in Einstein-Maxwell-Weyl gravity.
These solutions were separated in two groups according
to their seed solutions: Schwarzschild and non-Schwarzschild black hole solutions.
Since the Schwarzschild and the non-Schwarzschild black holes 'coalesce'
as $r_0=r_0^{min}\approx0.876$, we constructed the charged numerical black hole solutions
from both sides of this joint point. For each value of $r_0$,
there is a corresponding value $m_1=m_1^*$ that yields the
charged black hole solution as the increase of charge $Q$
without a singularity at spatial infinity. Later,
the thermodynamic proprieties of these charged black holes were
discussed in detail, and show that they obey the first law of thermodynamics.

Notice that the quasinormal modes of non-Schwarzschild solutions
have been investigated in refs.~\cite{Cai:2015fia,Zinhailo:2018ska},
which shows that the non-Schwarzschild black hole
is stable. Therefore it is necessary to investigate the
quasinormal modes and stability of the charged black hole
solutions in the Einstein-Maxwell-Weyl theory.
Another interesting possibility is (Anti-) de Sitter
charged black hole solutions in Einstein-Maxwell-Weyl gravity.
They have shown in the Einstein-Hilbert theory of gravity
with additional quadratic curvature terms \cite{Lin:2016kip}.
Beside the analytic Schwarzschild (Anti-) de Sitter solutions,
two groups of non-Schwarzschild (Anti-) de Sitter solutions
were also obtained numerically. Their thermodynamic properties
of the two groups of numerical solutions deserve a new work in future.

{\bf Acknowledgments}

We are grateful to Yun Soo Myung, Kai Lin, Hong L\"{u} and Yi-fu Cai for useful discussions.
This work was supported by National Natural Science Foundation of
China(Nos.11605152, 51575420 and 51802247).

\end{document}